\documentclass[manuscript]{aastex}
\usepackage{amsmath}

\begin{document}
\title{Solar parity issue with flux-transport dynamo}
\author{H. Hotta and T. Yokoyama}
\affil{Department of Earth and Planetary Science, University of Tokyo,
7-3-1 Hongo, Bunkyo-ku, Tokyo 113-0033, Japan}
\email{ hotta.h@eps.s.u-tokyo.ac.jp}

\begin{abstract} 
We investigated the dependence of the solar magnetic parity between the
 hemispheres on two important parameters, the turbulent diffusivity and
 the meridional flow, by means of axisymmetric kinematic dynamo
 simulations based on the flux-transport dynamo model. It is known that
 the coupling of the magnetic field between hemispheres due to turbulent
 diffusivity is an important factor for the solar parity issue, but the
 detailed criterion for the generation of the dipole field has not been
 investigated. Our conclusions are as follows. (1) The stronger
 diffusivity near the surface is more likely to cause the magnetic field
 to be a dipole. (2) The thinner layer of the strong diffusivity near
 the surface is also more apt to generate a dipolar magnetic field. (3)
 The faster meridional flow is more prone to cause the magnetic field to
 be a quadrupole, i.e., symmetric about the equator. These results show
 that turbulent diffusivity and meridional flow are crucial
 for the configuration of the solar global magnetic field.
\end{abstract}

\section{Introduction}
The solar magnetic $11$-year cycle is
 thought to be sustained by the dynamo motion of the
 internal ionized plasma \citep{1955ApJ...122..293P}. Based on 
the internal structure of the velocity
 field, i.e., the meridional flow and the differential rotation
revealed by helioseismology
 \citep[see review by][]{2003ARA&A..41..599T},
 the flux-transport dynamo was suggested
 \citep{1995A&A...303L..29C,1999ApJ...518..508D,2001A&A...374..301K,2010ApJ...709.1009H},
 as a model to
successfully explain some features of solar activity such as the
 equatorward
 migration  of sunspots and the poleward migration of the surface
 field.\par
 The solar global field has a distinct parity: the poloidal field is a
 dipole, i.e., antisymmetric
 about the equator. The polar fields almost always have the different
 sign between hemispheres, even though they show the occasional weak
 north-south asymmetry in phase and amplitude. In addition,
 Hale's polarity law states that
 the sunspots between hemispheres are nearly
 always antisymmetric about the equator
\citep{1908ApJ....28..315H}. It can then be interpreted that the toroidal
 fields ($B_\phi$) below the surface are antisymmetric about the equator.
 This interesting feature is, however, not axiomatically explained by
 the flux transport dynamo model since this model 
 significantly depends on three free parameters, i.e., the $\alpha$-effect, the
 meridional flow, and the turbulent diffusivity. \par
It has been suggested that the
 $\alpha$-effect around the base of the convection zone leads to the
 generation of the global dipolar
 magnetic field.
\citep{2001ApJ...559..428D,2002A&A...390..673B,2004A&A...427.1019C}.
The existence of the poloidal fields around the tachocline and the
 coupling of these fields between hemispheres are significant
 factors for the generation of the dipole field. A detailed
 explanation of this process is given in the next paragraph.
\cite{2004A&A...427.1019C} also suggested, however, that the dipole
 field can be obtained with the strong diffusivity in the convection
 zone without the presence of the the $\alpha$-effect around the
 tachocline.
 Hence the exact necessity of the $\alpha$-effect in generating the
 dipole field is still inconclusive.
\par
\begin{figure}
 \epsscale{0.9}
 \plotone{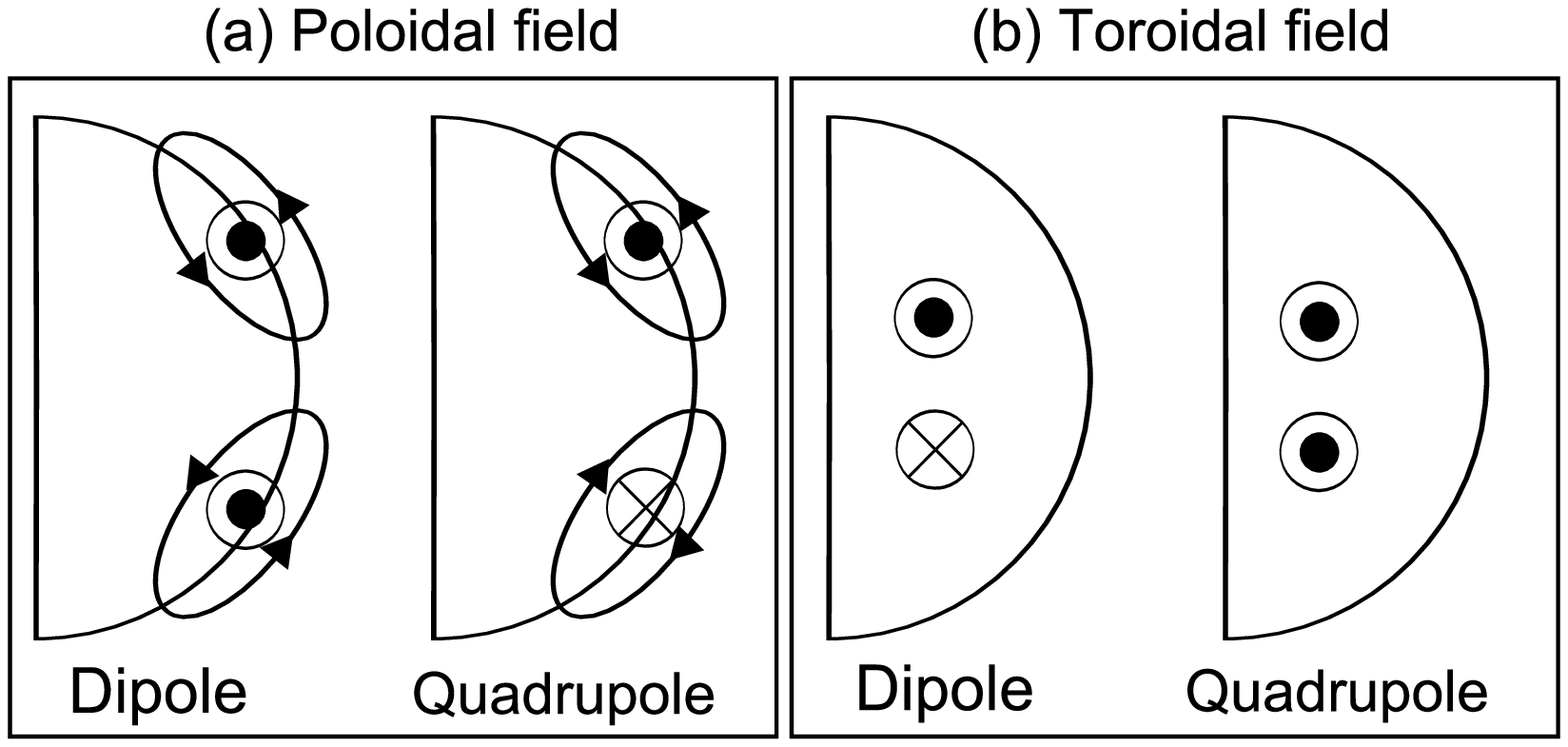}
\caption{Illustration of the parity issue.
Panel (a) shows the  poloidal fields (line) for a dipole and a quadrupole field and
 the  
corresponding vector potentials. Panel (b) shows the toroidal field for a
 dipole and a quadrupole. \label{explain}}
\end{figure}

The dependence of the parity on these parameters 
can be explained when we understand the role of the
 turbulent diffusivity in the
solar magnetic parity issue.
If the global magnetic field is antisymmetric, i.e. is a dipole like our
 sun, 
the $\phi$ component of the magnetic vector potential in each hemisphere
has the same sign (Fig. \ref{explain}a). When the cyclic phase in one hemisphere
slightly differs from the
other, the coupling effect by the turbulent diffusivity of the
poloidal field
distinguishes the phase difference in the vector potential and causes the
 magnetic field to be a dipole. On the other hand, when the
magnetic field is symmetric, i.e., is a quadrupole, this effect does
not occur. Therefore the substantial coupling of the poloidal field
generates the antisymmetric (a dipole) magnetic field.
The sign of the toroidal field in one hemisphere differs from that in the
other hemisphere when the global magnetic field is a solar-like dipole.
With the
same logic posited above, it is obvious that the
substantial diffusive
coupling of the toroidal field between the hemispheres helps the
 magnetic field to become
symmetric (a quadrupole; Fig \ref{explain}b).
In summary, the parity of the stellar global magnetic field depends on
which field, the toroidal or the poloidal, is more coupled
by the turbulent diffusivity between the hemispheres.
Detailed systematic parameter studies are necessary to understand for
 the parity issue.\par
In this study, we investigate the dependence of the solar magnetic
 parity on the
 distribution of the turbulent diffusivity and the amplitude of the
 meridional flow. 
The obtained constraint on the turbulent diffusivity is important since
 it is one of the key components of the solar dynamo model, although it is
 difficult to measure by direct observations.
\section{Model}\label{model}
We solve axisymmetric kinematic dynamo equations. The magnetic field
is divided into the toroidal field $B$ and the poloidal field
${\bf B_p}=\nabla\times [A(r,\theta){\bf e_\phi}]$
in the spherical coordinate $(r,\theta,\phi)$ as
\begin{eqnarray}
  {\bf B}=B(r,\theta){\bf e_\phi}+{\bf B_p}
\end{eqnarray}
where $A$ is the $\phi$ component of the magnetic vector potential and
${\bf e_\phi}$ is the unit vector along the $\phi$ direction.
Then the following standard forms of the dynamo equation are derived from
the magnetic induction equation as
 \begin{equation}
  \begin{split}
  \frac{\partial B}{\partial t}&+\frac{1}{ r}
  \left[
   \frac{\partial}{\partial
   r}(ru_rB)+\frac{\partial}{\partial\theta}(u_\theta B)
  \right] \\
 &=r\sin\theta({\bf B_p}\cdot\nabla)\Omega 
 -(\nabla\eta\times\nabla\times
  B{\bf
  {e}_\phi})\cdot{\bf e_\phi} \\
&+\eta
  \left(
   \nabla^2-\frac{1}{r^2\sin^2\theta}
  \right)B,\label{eq:dynamo1}
\end{split}
\end{equation}
\begin{equation}
 \begin{split}
\frac{\partial A}{\partial t}&+\frac{1}{r\sin\theta}({\bf
 u}\cdot\nabla)(r\sin\theta A)\\
&=\eta
  \left(
   \nabla^2-\frac{1}{r^2\sin^2\theta}
  \right)A
+S(r,\theta;B).\label{eq:dynamo2}
 \end{split}
\end{equation}
We specify the meridional flow ${\bf u}=u_r{\bf e_r}+u_\theta{\bf
e_\theta}$, the differential rotation $\Omega$, and the turbulent 
diffusivity $\eta$.
A source term $S(r,\theta;B)$ is artificially added to the 
right-hand side of equation (\ref{eq:dynamo2}).
This source term describes the generation of poloidal fields at the solar
surface from the decay of bipolar sunspots. This is the
``Babcock-Leighton $\alpha$-effect''
\citep{1969ApJ...156....1L,1961ApJ...133..572B}.
Once these quantities are specified, we
can solve equations (\ref{eq:dynamo1}) and (\ref{eq:dynamo2}) to study the
behavior of the dynamo kinematically.\par
The  formula for the differential rotation is given as
\begin{equation}\label{omega}
 \Omega(r,\theta)=\Omega_c+\frac{1}{2}
\left[
1+\mathrm{erf}
\left(
2\frac{r-r_c}{d_c}
\right)
\right]\{\Omega_s(\theta)-\Omega_c\}
\end{equation}
where $\Omega_s(\theta)=\Omega_{\mathrm{Eq}}+a_2\cos^2\theta+a_4\cos^4\theta$
is the surface latitudinal differential rotation and
$\mathrm{erf}$ is the error function. The parameters are
set as $\Omega_c/2\pi=432.8\ \mathrm{nHz}$,
 $\Omega_{\mathrm{Eq}}/2\pi=460.7\ \mathrm{nHz}$, $a_2/2\pi=-62.69
\ \mathrm{nHz}$, $a_4/2\pi= -67.13\ \mathrm{nHz}$,
 and $r_c=0.7R$ and $d_c=0.05R$, which closely resemble the best-fit
helioseismic solution. $R$ is the solar radius.
$r_c$ and $d_c$ is the central radius and the thickness of the
tachocline, respectively. $\Omega_c$ is the rotating rate of the core.
This differential rotation profile has a purely
latitudinal difference with equatorial acceleration in the convection 
zone. It smoothly matches across the ``tachocline'' with the
core rotating rigidly. \par
We now describe how the meridional flow is specified. While a poleward
meridional flow is observed near the solar surface, the structure of the
internal return flow is at present unconstrained observationally.
We choose here an analytical form suggested by \cite{1988ApJ...333..965V}
 with the density profile in the convection zone given 
by $\rho(r)\propto [(R/r)-1]^m$; the $r$ and $\theta$ components of this
flow are as follows:
\begin{equation}
\begin{split}
u_r&(r,\theta)=
\frac{u_0}{f}
\left(\frac{R}{r}\right)^2\\
&\times
\left[
-\frac{1}{m+1}+\frac{c_1}{2m+1}\xi^m
- \frac{c_2}{2m+p+1}\xi^{m+p}
\right]\\
& \times\xi\sin^q\theta[(q+2)\cos^2\theta-\sin^2\theta]\label{meridional1}
\end{split}
\end{equation}
\begin{equation}
\begin{split}
u_\theta(r,\theta)
=&
\frac{u_0}{f}
\left(\frac{R}{r}\right)^3
[-1+c_1\xi^m-c_2\xi^{m+p}]\\
&\times\sin^{q+1}\theta\cos\theta,\label{meridional2}
\end{split}
\end{equation}
where
\begin{eqnarray}
 \xi(r)=\frac{R}{r}-1,
\end{eqnarray}
\begin{equation}
 c_1=\frac{(2m+1)(m+p)}{(m+1)p}\xi^{-m}_p,
\end{equation}
\begin{eqnarray}
 c_2=\frac{(2m+p+1)m}{(m+1)p}\xi^{-(m+p)}_p,
\end{eqnarray}
\begin{eqnarray}
 \xi_p=\frac{R}{r_p}-1. 
\end{eqnarray}
As shown in eqs. (\ref{meridional1}) and (\ref{meridional2}),
$u_0$ is the amplitude of the velocity and $p$ and $q$ are respectively
the radial and
latitudinal dependence of the flow.
We specify the bottom of the meridional flow $r_p$, and the
normalization constant $f$
to set the 
maximum speed of the meridional flow of the $\theta$ component to
$u_0$.
 We use the parameter values $m=0.5$, $p=0.25$, $q=0$ and $r_p=0.62R$. 
Note that our meridional flow slightly penetrates into the rigidly
rotating core, i.e. $r_p<r_c$ \citep{2010ApJ...709.1009H}.\par
We assume that the net magnetic diffusivity in the convection zone is dominated
by its turbulent contribution. We adopt a diffusivity profile of the form 
\begin{equation}
\begin{split}
 \eta(r)=&\eta_{\mathrm{core}}+\frac{\eta_t}{2}
\left[
1+\mathrm{erf}\left(\frac{r-r_1}{d_1}\right)
\right]\\
&+\frac{\eta_{\mathrm{s}}}{2}
\left[
1+\mathrm{erf}\left(\frac{r-r_2}{d_2}\right)
\right].
\end{split}
\end{equation}
Here, $r_1=0.7R$, $d_1=0.02R$ and $d_2=0.02R$.
This profile consists of three layers.
In the strong diffusivity layer $(r>r_2)$,
the diffusivity is prescribed by $\eta_{\mathrm{s}}$ within
$10^{12}-10^{14}\ \mathrm{cm^2}\
\mathrm{s^{-1}}$. \cite{1989ApJ...347..529W}
 argued
that the surface diffusivity should be $6\times10^{12}\ \mathrm{cm^2}\
\mathrm{s^{-1}}$ to be consistent with observations of the time
development of the surface magnetic field.
In the convection zone we use the fixed value
 $\eta_t=5\times10^{10}\ \mathrm{cm}^2\ \mathrm{s}^{-1}$.
In the subadiabatically stratified core there is no turbulence (or at
least, far less), so that the diffusivity is presumably much weaker.
We use the value 
$\eta_{\mathrm{core}}=5\times10^8\ \mathrm{cm}^2\ \mathrm{s}^{-1}$.
For convenience, we define the surface depth $d_s=R-r_2$ which denotes
the thickness of the strong diffusivity layer.
We take $d_s$ and $\eta_\mathrm{s}$ as free parameters, since they affect
the parity of the magnetic field. \par
Some parts of the toroidal field in the tachocline rise to the surface 
due to the magnetic buoyancy and generate active regions.
During this process, the flux tube expands and the Coriolis force bends
the flux.
There is observational evidence offered by \cite{1959ApJ...130..364B} that the 
decay of tilted bipolar active regions can produce a substantial
amount of the net poloidal fields near the surface \citep{1991ApJ...375..761W}.
This is called the ``Babcock-Leighton $\alpha$-effect.''
Based on this, we assume that the poloidal source term is taken in the form
\begin{equation}
\begin{split}
 S(r,\theta;B)=&\alpha(r,\theta)B(r_c,\theta)\\
&\times\left[  \frac{1}{1+(B(r_c,\theta)/B_{eq})^2} \right],
\label{alpha1}
\end{split}
\end{equation}
where
\begin{equation}
\begin{split}
 \alpha(r,\theta)=&\frac{s_0}{4}\\
&\times\left[
1+\mathrm{erf}\left(\frac{r-r_4}{d_4}\right)
\right]\left[
1-\mathrm{erf}\left(\frac{r-r_5}{d_5}\right)
\right]\\
&\times
\sin\theta\cos\theta\left[\frac{1}{1+e^{-\gamma(\theta-\pi/4)}}\right].
\label{alpha}
\end{split}
\end{equation}
The parameters are $r_4=0.95R$, $r_5=R$, $d_4=0.05R$, $d_5=0.01R$,
 and $\gamma=30$.
 We concentrate the $\alpha$-effect by the last factor in eq. (\ref{alpha})
at the low latitude in which there are
 many observed active regions \citep{2004ApJ...601.1136D}.
The source term is made proportional to the toroidal
field strength at the same latitude in the tachocline $r=r_c$
 (eq. (\ref{omega})), i.e. the base
of the convection zone, since it is assumed here to originate from
 the radially emerged magnetic fluxes.
The quenching term $\{1+[B(r_c,\theta)/B_{\mathrm{eq}}]^2\}^{-1}$
 in eq. (\ref{alpha1}) ensures that 
the poloidal field production rapidly vanishes as the deep toroidal field 
strength exceeds $B_{\mathrm{eq}}$
\citep{PhysRevE.54.R4532}. $B_{\mathrm{eq}}$ is the equipartition
 magnetic field.
The Coriolis force
 cannot bend a strong magnetic field ($>B_\mathrm{eq}$) in the convection zone.
We use a fixed value $B_{\mathrm{eq}}=4\times10^4\ \mathrm{G}$.
$s_0$ is determined by fixing the value of the dynamo number (see \S
 \ref{result} in detail).
\par
We solve equations (\ref{eq:dynamo1}) and (\ref{eq:dynamo2}) numerically
in 
all the sphere of the
meridional plane in $0.6R<r<R$ and $0<\theta<\pi$ with the modified
Lax-Wendroff scheme.
We use a moderate resolution of around 64 grid points in the radial
direction and 128 grid points in the latitudinal direction.
 At the lower boundary ($r=0.6R$), we set both $B$
and $A$ at zero,
indicating that the radiative core is a perfect conductor.
At the top boundary ($r=R$), we set $B=0$ and smoothly match $A$ onto
an exterior potential field solution \citep{1994A&A...291..975D}.
 At both poles ($\theta=0$ and $\pi$), we set $B=A=0$ for the regularity.
The numerical convergence is checked by runs with different
grid spacings.
\section{Results}\label{result}
A new indicator of the magnetic parity is defined in this study.
The radial magnetic field at the surface can be decomposed as
\begin{eqnarray}
 B_r(R,\theta)=\sum_{n=0}c_nP_n(\cos\theta),
\end{eqnarray}
where $P_n$ is the Legendre polynomial. Then we define the symmetric parameter as
\begin{eqnarray}
 \mathrm{SP}=\frac{\displaystyle\sum_{i=0}|c_{2i}|-\sum_{i=0}|c_{2i+1}|}
  {\displaystyle\sum_{\mathrm{i=0}}|c_i|}.
\end{eqnarray}
Each even (odd) order of the Legendre polynomial is symmetric (antisymmetric)
about the equator.
Therefore, $\mathrm{SP}=1$ corresponds to the purely symmetric mode about the equator and
$\mathrm{SP}=-1$ is the antisymmetric mode. \par
\begin{figure}
 \epsscale{0.9}
 \plotone{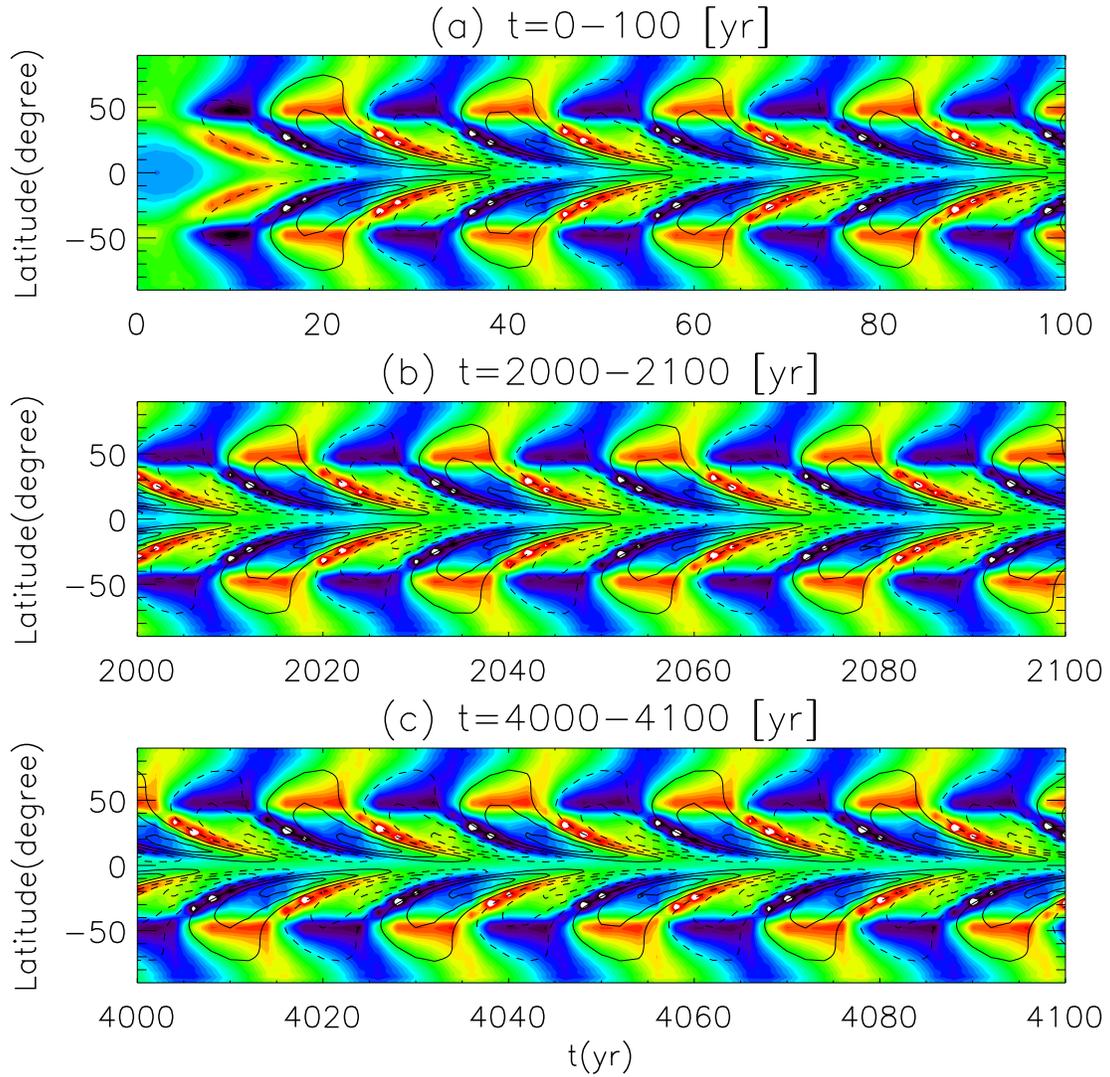}
\caption{Butterfly diagram for the reference solution. Time-latitude
 plot of $B_\phi|_{r=0.7R}$ by contour is superposed on the color map of the
 surface radial fields.\label{butterfly}}
\end{figure}
\begin{figure}
 \epsscale{0.9}
 \plotone{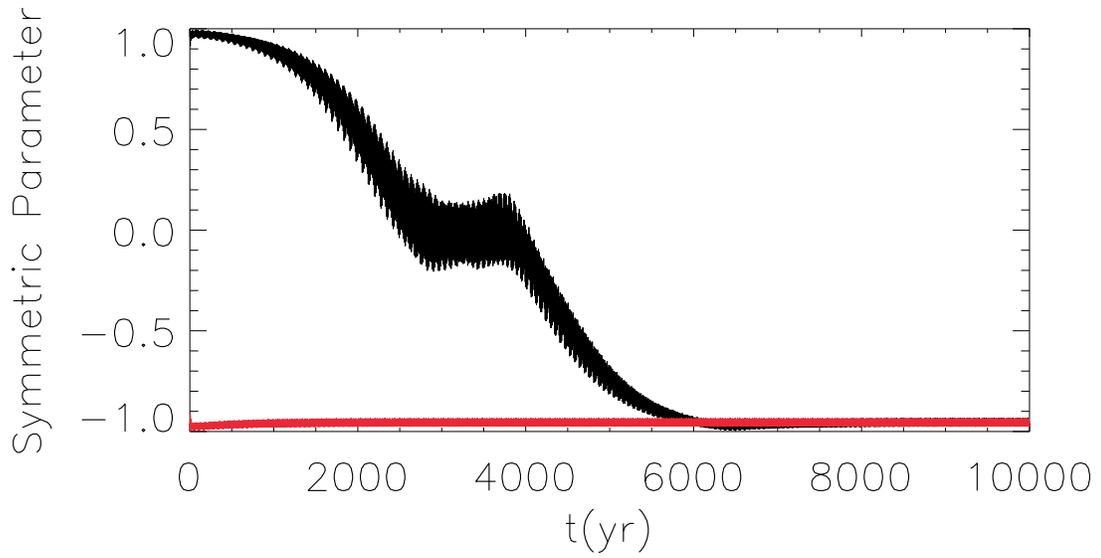}
\caption{Time-development of the symmetric parameters.
The black (red) line corresponds to the results of the
 symmetric (antisymmetric) initial
 condition. Regardless of the initial condition the symmetric parameter
 finally becomes $\sim -1$ (antisymmetric solution).\label{sym}}
\end{figure}

We first show a representative reference solution, computed with
the amplitude
of the meridional flow $u_0=1000\ \mathrm{cm}\ \mathrm{s^{-1}}$,
the amplitude of the $\alpha$-effect $s_0=160\ \mathrm{cm\ s^{-1}}$,
the amplitude of the turbulent diffusivity near the surface layer
$\eta_{\mathrm{s}}=2\times10^{12}\ \mathrm{cm^2}\ \mathrm{s^{-1}}$
and 
the thickness of the strong diffusivity layer $d_\mathrm{s}=0.1R$.
The results are shown in the time-latitude plots in
Fig. \ref{butterfly}. This simulation is started with a symmetric
initial condition. As time passes, the global magnetic field becomes
antisymmetric. In such a case, the symmetric parameter develops as
shown in Fig. \ref{sym}. The black (red) line denotes the result with a
symmetric (antisymmetric) initial condition. Regardless of the initial
condition, the parity of the magnetic field approaches the stationary
antisymmetric state where
 the symmetric parameter becomes $\sim -1$.
We conclude that the magnetic field with these reference parameters
finally becomes a dipole field.\par
We investigate the asymptotic stationary values of the symmetric
parameter for runs in different setups.
We carried out runs by choosing a value for the surface diffusivity
$\eta_\mathrm{s}$, from 8
points in the range $6\times10^{11}$ to
$1\times10^{13}$ $\mathrm{cm^2\ s^{-1}}$ 
and the surface depth $d_\mathrm{s}$, from 5 points in the range $0.1R$
to $0.25R$. We specify the amplitude of the $\alpha$-effect by
\begin{eqnarray}
 s_0=160\ \mathrm{cm\ s^{-1}}
\left(
\frac{\eta_\mathrm{s}}{2\times10^{12}\ \mathrm{cm^2\ s^{-1}}}
\right)^2
\left(
\frac{d_\mathrm{s}}{0.1R}
\right)^2
\end{eqnarray}
The background reason for this setup is to fix the value of the surface
integrated
dynamo number,
\begin{eqnarray}
 N_D=\frac{s_0 k_x R d_\alpha}{\eta^2_s d_s^2k^4}\frac{d}{dr}
(r\Omega\sin\theta),
\end{eqnarray}
where $k_x$ and $k$ denote the wavenumber of the magnetic field
and $d_\alpha$ denotes the thickness of the layer where the
$\alpha$-effect
 is effective, and where $k_x$, $k$ and $d_\alpha$ are assumed to
be unchanged from case to case.
 This dynamo number is kept unchanged since it determines the oscillatory
 nature of the dynamo.
The reason for this idea is because the strong $\alpha$-effect is necessary
to endure 
the diffusivity in the large area.
We also vary the amplitude of the meridional flow:
$u_0=1000\ \mathrm{cm\ s^{-1}}$(slow meridional flow case)
 and $2000\ \mathrm{cm\ s^{-1}}$(fast meridional flow case). For
every parameter set, we conducted runs with both symmetric and antisymmetric
initial conditions to ensure that the asymptotic value of the symmetric
parameter does not depend on the initial parity. 
$8$(diffusivity)$\times4$(surface depth)$\times2$(meridional
flow)$\times2$(initial parity)$=128$ simulation runs carried out.
All simulations are calculated for more than $10000$ years.
\par
\begin{figure}
 \epsscale{0.9}
 \plotone{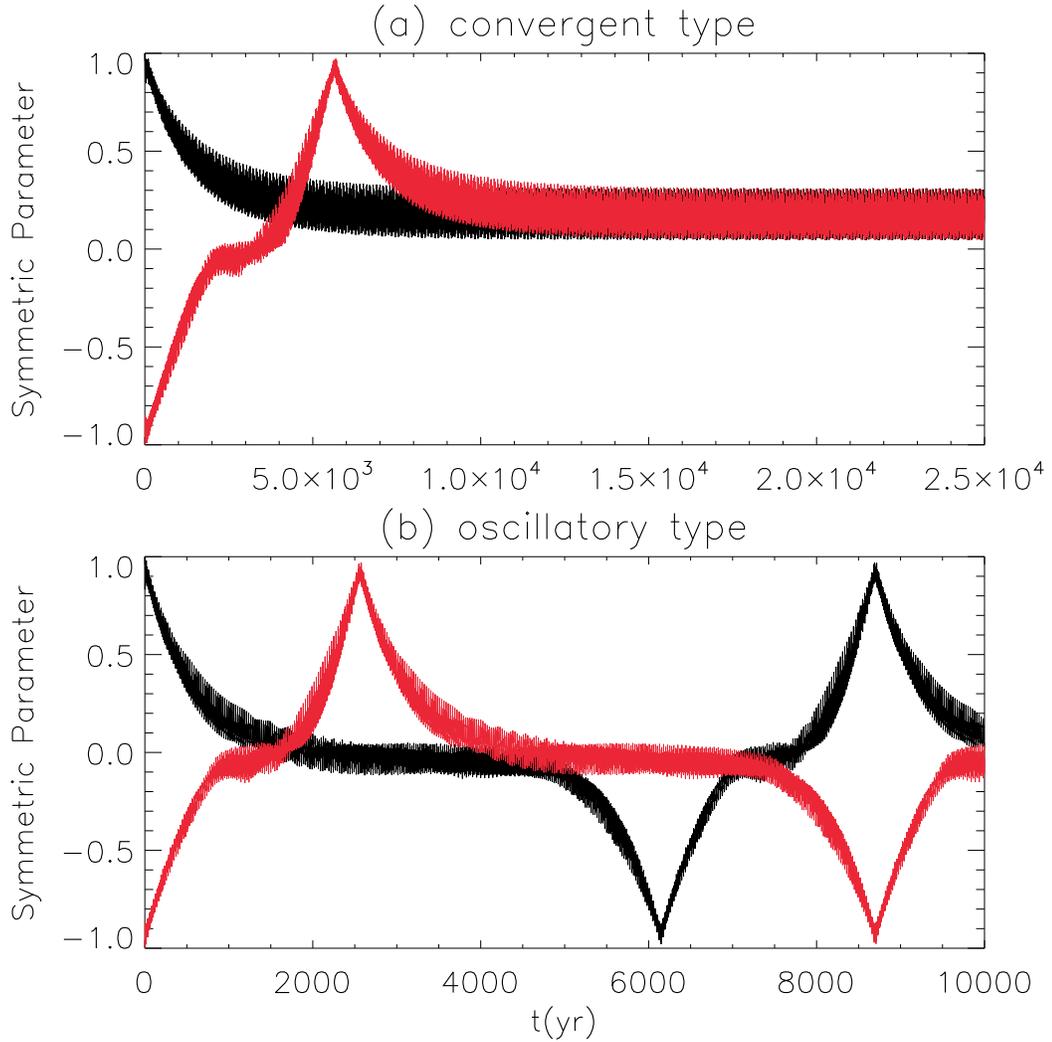}
\caption{
 Two types in "zero" symmetric parameter cases.
The time-development of the symmetric parameter for each type is
shown. The format is the same as in Fig. \ref{sym}. Panel (a) shows the convergent
type in which the value of the symmetric
parameter finally converges to zero. Panel (b) shows the oscillation type in
which the symmetric parameter continues to
oscillate between the quadrupole (SP$\sim -1$) and the dipole (SP$\sim 1$) solutions.
\label{mix}}

\end{figure}

There are two types of solutions when the value of the SP is around zero
and we categorize them as ``mixed-partity'' cases.
One type is similar to the reference case (Fig. \ref{mix}a). The value of the SP
finally converges. The other type is interesting in that the value of the
symmetric parameter does not converge and continues 
to oscillate between the quadrupole and the dipole solutions (Fig. \ref{mix}b).
Since the averaged value in the calculation duration is close to zero,
we adopted it for the SP in such cases. 
\par
\begin{figure}
 \epsscale{0.9}
 \plotone{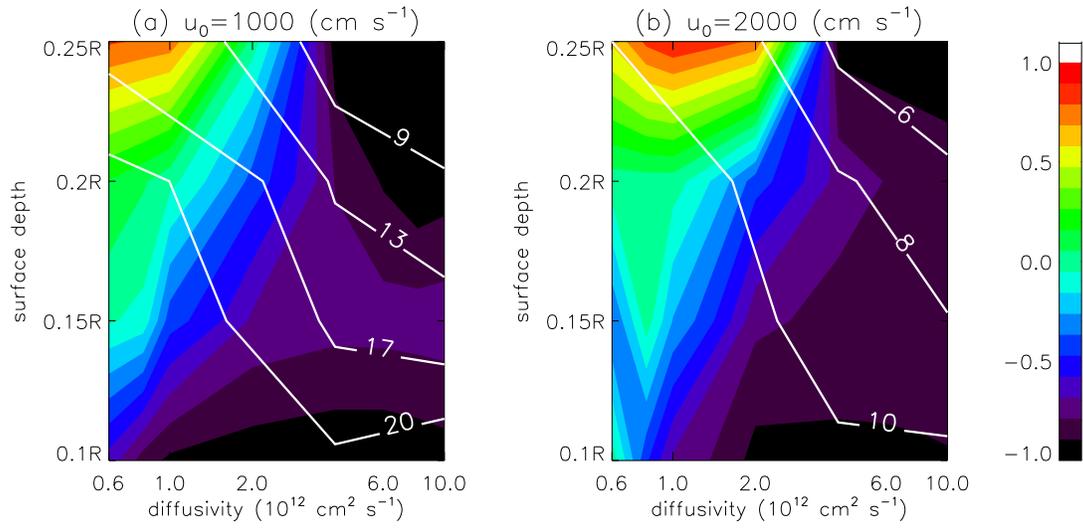}
\caption{
Symmetric parameter SP as a function of the diffusivity
 $\eta_\mathrm{s}$ and the surface depth $d_\mathrm{s}$.
The superposed lines indicate the contours of the dynamo cycle
 period over periods of years.
Panel (a) shows the results for the slow meridional flow case 
($u_0=1000\ \mathrm{cm\ s^{-1}}$).
Panel (b) shows the result for the fast meridional flow case 
($u_0=2000\ \mathrm{cm\ s^{-1}}$). \label{pam}}

\end{figure}

The results of this parameter space study are shown in Fig. \ref{pam}.
The dynamo cycle period is also shown by the contour lines. The period is
shorter when the surface depth is thicker since the transport of the magnetic
flux by the diffusivity is more effective.
Panel (a) shows the result of the slow meridional flow case
($u_0=1000\ \mathrm{cm\ s^{-1}}$). Two points can be ascertained
from this figure.
One is that regardless of the surface depth, the
strong diffusivity ($> 3\times10^{12}\mathrm{\ cm^2}\ s^{-1}$) can make
the magnetic
field to become a dipole (SP$\sim-1$). The other is that with the thinner
surface depth, no strong diffusivity ($>
1\times10^{12}\mathrm{\ cm^2}\ s^{-1}$) is needed to generate the dipole field.
This means that the magnetic field is more likely to be a dipole with
the thinner surface depth.
Fig. 4b shows the result of the fast meridional flow case ($u_0=2000\
\mathrm{cm\ s^{-1}}$). It is obvious that the parameter area for the
symmetric solutions, i.e. $S_\mathrm{p}>0$,  increases. This indicates
that the fast
meridional flow causes the magnetic field to be symmetric.
\section{Discussion and Conclusion}\label{conclusion}
We investigated the dependence of the global magnetic parity on the
distribution of the diffusivity (the amplitude and the surface depth)
and the amplitude of the meridional flow.
Three results were obtained. First, the model shows that the stronger
diffusivity near the surface
acts to make the
magnetic field a dipole. The diffusivity near the surface
enhances mainly the coupling of the poloidal field near the surface
between the hemispheres, leading to the generation of dipolar magnetic field.
The second result is that the thinner layer of
the strong surface diffusivity also works to cause the magnetic field to become
dipolar. The thinner surface depth suppresses the coupling of the toroidal
field between the hemispheres since most of the toroidal field exists
around the tachocline.
The third result is that the fast meridional flow causes the
magnetic field to become a quadrupole. The fast meridional flow prevents the
poloidal field from coupling  near the surface of the equator
because the flow
transports the poloidal field poleward. In addition, the flow transports
the toroidal field around the tachocline equatorward, thus causing
the coupling of the toroidal field. These three results quantitatively constrain the
distribution and the amplitude of turbulent diffusivity, which cannot be
determined by observation and is a important factor for the dynamo problem.
\par
In this study, we did not investigate the dependence of the parity on the
$\alpha$-effect in the convection
zone, which may be a strong factor in causing the magnetic field
to become a dipole. The
poloidal field generated by this effect
 around the tachocline is transported equatorward by the
meridional flow, and this process enhances the coupling of the poloidal field
between the hemispheres
\citep{2001ApJ...559..428D,2002A&A...390..673B,2004A&A...427.1019C}.
It is possible that the criterion for a dipole field we obtain in this
study may be modified with this type of $\alpha$-effect.
We will discuss the possibility of the existence and the influence of
the $\alpha$-effect in a forthcoming paper.
Another interesting issue to be addressed is the
possibility that the variation of the velocity
field in the solar cycle affects the parity. 
In the calculations for the earth dynamo there is the significant difference
between the kinematic and the MHD cases in the parity issue
\citep{nishikawa2008simulation}.
Thus, in the future we will 
investigate the parity issue with the Lorentz feedback
\citep{2006ApJ...647..662R}. 

\acknowledgments
Numerical computations were carried out on the general-purpose PC farm
at the Center for Computational Astrophysics (CfCA) of the National Astronomical
 Observatory of Japan. 
The page charge for this paper is supported by CfCA.
We have greatly benefited from the proofreading/editing assistance from the GCOE program.

\end{document}